\documentclass[notoc]{JHEP}
\usepackage{amssymb}
\def\R{{\mathbb R}}
\def\M{{\mathbb M}}
\def\P{{\mathbb P}}
\def\Z{{\mathbb Z}}
\def\Cl{{\mathbb C}\ell}
\def\be{\begin{equation}}
\def\ee{\end{equation}}
\def\bea{\begin{eqnarray}}
\def\eea{\end{eqnarray}}
\title{On conformal reflections in compactified phase space}
\author{P. Budinich \\International School for Advances Studies and \\ 
International Center for Theoretical Physics, \\ Strada Costiera 13, 34014 Trieste,Italy}

{\abstract{
Some results from arguments of research dealt with R. Raczka are exposed
and extended. In particular new arguments are brought in favor of the
conjecture, formulated with him, that both space-time and momentum may be
conformally compactified, building up a compact phase space of
automorphism for the conformal group, where conformal reflections
determine a convolution between space-time and momentum space which may
have consequences of interest for both classical and quantum physics.
}}

\begin{document}
\setcounter{page}{1}
\section{Introduction}

In 1931 P.M.A. Dirac wrote \cite{1}: 
{\it ``There are at present fundamental problems
in theoretical physics ... whose solution will presumably require a more
drastic revision of our fundamental concepts than any that have gone
before. Quite likely these changes will be so great that it will be beyond
the power of human intelligence to get the necessary new ideas by direct
attempts to formulate the experimental data in mathematical terms. The
theoretical work in the future will therefore have to proceed in a more
indirect way. The most powerful method of advance that can be suggested at
present is to employ all the resources of pure mathematics in attempts to
perfect and generalize the mathematical formalism that forms the existing
basis of theoretical physics, and after each success in this direction, to
try to interpret the new mathematical features in terms of physical
entities ...''}. Later he added: {\it ``It seems that, if one is working from 
the point of view of getting beauty in one's equations, and if one has really
sound insight, one is on a sure line of progress''}.

Dirac started himself the process of research in abstract mathematics,
aimed at the understanding of physical quantum phenomena, which brought him
to discover his beautiful spinor field equation which, not only explained
the
origin of electron spin, but even anticipated the discovery of new,
unexpected physical phenomena; as those deriving from the existence of
antimatter. He was certainly a forerunner and now, after 67 years, when
looking at the recent developments in theoretical physics, where some
branches like quantum groups, noncommutative geometry, deal mainly with
subjects of pure mathematics, his words sound like a prophecy. 

Having in mind his recommendations, an undeniable source of beauty, of
relevance for quantum physical phenomena, may certainly be found in the E.
Cartan work on simple spinor geometry \cite{2}. 
E. Cartan stressed specially two
concepts.
\begin{enumerate}
\item Vectors of euclidean spaces may be conceived as bilinearly composed
by
spinors.
\item Rotations in vector spaces may be decomposed in products of reflections.
\end{enumerate}

Notoriously, the main geometrical tool for the description of the phenomena
of classical mechanics is euclidean geometry in pseudo euclidean vector
spaces and rotation therein. Following E. Cartan, spinors and reflection
may be conceived as the elementary constituents of these. 

The study of covariance of the equations of motion with respect to rotation
groups has been one of the main mathematical instrument of research in
classical mechanics in the last and also in the present century, when
covariance of Maxwell's equations with respect to the Lorentz-Poincar\'e
groups has brought Einstein to the discovery of special relativity and of
the geometrical structure of space-time: $\M=\R^{3,1}$. 

The importance of reflection groups, instead, has been only recognized
after the advent of quantum mechanics. In fact space-time reflection group play
an important role in the explanation of quantum phenomena. Precisely space
reflections allow us to define the concept of parity (not conserved in weak
decays) while time inversion allows us to understand the existence of
antimatter. The most appropriate space for dealing with such reflections 
seems to be spinor space.

It is known since 1909 that Maxwell's equations are also covariant with
respect to the conformal group. A discovery which has also brought to the
conjecture that Minkowski space-time may be conformally compactified, and
represented by a particular realization of an homogeneous space of the
conformal group. With R. Raczka, somehow in line with Dirac's
recommendations, we have observed that another realization of that
homogeneous space could well represent conformally compactified momentum
space which, together with compactified space-time, could then build up a
compact phase space where both the concept of infinity and that of
infinitesimal, of difficult, if not impossible, self consistent mathematical
definition, would not be needed, allowing then not only a rigorous
mathematical formulation of theoretical physics, but also the elimination
of the main difficulties encountered by quantum physics: the ones of so
called infrared and ultraviolet divergences.

The conformal group, in which the Lorentz-Poincar{\'e} group is contained
as a 
subgroup, has a reflection group which contains, beside space-time
reflections, mentioned above, also two additional reflections, called conformal
reflections. However, somehow surprisingly, despite the relevance of
conformal covariance for the understanding of several physical phenomena concerning massless systems,
these reflections have failed, up to now, to manifest their role in
physics. 

One of purposes of this paper is to review some of the results obtained in
the work with R. Raczka and to outline some follow up of our thinking,
 in particular on the possible role in physics of conformal reflections,
obtained after he left us.

\section{Compact phase space.}

In a similar way as, from Maxwell's equations Lorentz covariance, Minkowski
derived the pseudo euclidean structure of space-time: $\M = \R^{3,1}$, 
from their
conformal covariance Veblen derived and adopted \cite{3} the conformally 
compactified structure $\M_c$ of $\M$:
\be
 \M_c=\frac{S^3 \times S^1}{\Z_2} \;\; ,
\label{1}
\ee

in which Minkowski space-time $\M$ is densely contained (which means: to every
point of $\M_c$ there correspond one point of $\M$; to a submanifold of $\M_c$,
 of dimension 3, there corresponds the points of $\M$ at infinity) afterwards
also adopted by several authors \cite{4}. 

We will show now that the same argument which induces to postulate that $\M$
is compactified induces also to postulate that its Fourier dual momentum
space $\P=\R^{3,1}$ is conformally compactified as well.

In fact the conformal group may be linearly represented by $O(4,2)$ acting in
$\R^{4,2}$. Let us consider the  equations for the corresponding Weyl spinors 
or twistors:
\be
 \sum_{a=1}^6 p_a \gamma^a \left(1 \pm \gamma_7 \right) \pi = 0 \;\; ,
\label{2}
\ee

where $p_a \in  \R^{4,2}$, $\gamma^a$  are the generators of the 
Clifford algebra
$\Cl(4,2)$, $\gamma_7$ its volume elements and $\pi$ is a vector of the 
spinor space $S$ defined by ${\rm{End} }S =\Cl(4,2)$. 

Eq. (\ref{2}) for non zero twistors 
$\pi_\pm=\frac{1}{2}\left(1\pm \gamma_7\right)\pi$, implies $p_ap^a = 0$, 
and therefore the directions of $p_a$ form the projective quadric $\P_c$:
\be
 \P_c = \frac{S^3 \times S^1}{\Z_2}.
\label{3}
\ee

It may be easily seen that eq. (\ref{2}) contains the equations:
\be
  \sum_{\mu=0}^3 p_\mu \gamma^\mu \varphi_\pm=0 \;\;\; ,
\label{5}
\ee

where $\varphi_\pm$ represent Weyl spinors associated with $\Cl(3,1)$, 
of which $\gamma^\mu$ are the generators. Eq. (\ref{5}) is the Weyl equation 
in momentum space $\P=\R^{3,1}$ ,
Fourier dual of Minkowski space $\M=\R^{3,1}$, from it is easy to obtain 
\cite{5} the equations:
\be
 p_\mu F_+^{\mu\nu}=0 \; ; \;\;\;\;\; 
  p_\mu F^{\mu\nu}_-=0 \;\; ,
\label{6}
\ee

where
$$
 F_\pm^{\mu\nu}=\langle \varphi_\pm^\dagger 
  \left[ \gamma^\mu,\gamma^\nu \right] \varphi_\pm \rangle \;\; .
$$

Eq.s (\ref{6}) represent the homogeneous Maxwell's equations, 
in momentum space $\P=\R^{3,1}$,
which then results densely contained in $\P_c$ of eq. (\ref{3}).

This short-cut derivation of Maxwell's equations in momentum space from
twistors equations indicates that, in so far, Maxwell's equations conformal
covariance implies the conformal compactification $\M_c$, as given in 
eq. (\ref{1}), of Minkowski space $\M$, their derivability from twistors 
equations in momentum
space implies the conformal compactification $\P_c$ of momentum space $\P$, 
as given in eq. (\ref{3}), as well.

The resulting phase space will  then be compact and consequently any field theory
formulated in such a compact phase space should, a priori, expected to be
free from both infrared and ultraviolet divergences.

The main problem will be to define, for every function $f(x)$ taking values
in $\M_c$, a transform to a function $F(k)$, taking values in $\P_c$, such 
that in the flat limit (radiuses of $S^3$ and $S^1$ going to infinity) it 
identifies with the standard Fourier transform correlating $\M$ and its 
dual $\P$.

 The problem may be solved \cite{6} for the two-dimensional case 
$\M =\R^{1,1}= \P$, for which:
\be
 \M_c = \frac{S^1 \times S^1}{\Z_2}=\P_c \;\;\;  ;
\label{7}
\ee	

one needs only to inscribe in each $S^1$ a regular polygon with 
$2 N = 2 \pi R K $
vertices, where $R$ (of dimension $[\ell]$: length) and $K$ (of dimension 
$[\ell^{-1}]$ are
 the radiuses of the $\M_c$ and $\P_c$ circles respectively. They define in 
$\M_c$ and $\P_c$ two lattices: $M_L \subset \M_c$; $P_L \subset \P_c$ which are Fourier 
dual. Indicating in
fact with $f(x_{nm})$ a function taking values in $M_L$ and with  
$F(k_{\rho\tau})$ a
function taking values in $P_L$ we have : 
\bea
 f\left(x_{nm}\right)=\frac{1}{2 \pi R^2} \sum_{\rho, \tau = -N}^{N-1} 
  \varepsilon^{\left(n \rho - m \tau\right)} F\left(k_{\rho \tau}\right) 
\;\; , \nonumber \\
\rm{~~~~} \;\;\;\;\; \label{8} \\
 F\left( k_{\rho\tau} \right) = \frac{1}{2 \pi K^2} \sum_{n,m = -N}^{N-1}
  \varepsilon^{-\left(n \rho - m \tau \right)} f \left(x_{nm}\right) 
\;\; , \nonumber
\eea

	where  $ \varepsilon=e^{i \frac{\pi}{N}}$ is the $2N$-root of unity.

	Eq.s (\ref{8}), in the limit $R, K \rightarrow \infty$ may be easily 
identified with the standard
Fourier transforms in $\M=\R^{1,1}=\P$.

	It is obvious that in $M_L$ and $P_L$ any field theory will be free 
from both infrared and ultraviolet divergences.

	In the realistic, four dimensional case, since, in principle, 
the concept
of infinity and infinitesimal should not be realizable, one could expect
again that phase space should restrict to discrete and fine lattices which 
however do not seem to be obtainable with standard mathematical algorithms 
\cite{7}, through
which, instead one may anticipate some aspects of the convergences of field
theories in conformally compactified space-time and momentum space.

In fact it is known that the space $\P_c$, given by eq. (\ref{3}), is 
conformally flat i.e
$$
 g_{\mu\nu}\left(p\right) = \Omega^2\left(p\right) \eta_{\mu\nu},
$$
where $\Omega\left(p\right)$ is the conformal factor and $\eta_{\mu\nu}$ is 
the metric tensor of flat $\R^{3,1}$. $\Omega\left(p\right)$   	 
may be obtained \footnote{
 It is interesting to observe that $\Omega(p)$ may be rigorously
 set  in the form $\Omega(p)=M^2/p^2_W$, where $p^2_W$ is the Wick rotated
$p^2$, that is
$p^2_W=p_1^2+p_2^2+p_3^2+p_0^2$ which ensures the non singularity of 
 $\Omega(p)$, which instead is not guaranteed by (\ref{10}) and (\ref{10a}) for
$p^2$ time-like and space-like  respectively.
} 
by adopting the Dirac six-dimensional formalism:
 $p_\mu=P_\mu/\left(P_5+P_6\right)$ and, as shown in reference \cite{8}, 
it provides a convergence factor since:
\be
 \lim_{|p|^2 \rightarrow \infty} \Omega^2(p) = \frac{M^4}{p^4} \;\; ,
\label{9}
\ee
	
 where $M$ is a mass scale ($c=1$).

\noindent In particular for the de Sitter subgroup $SO(4,1)$ (obtained for 
$P_6 = 1$):
\be
 \Omega\left(p\right)=\frac{4 M^2}{4M^2+p^2} \;\; ,
\label{10}
\ee

identical to the Pauli-Villars regularizing factor, often adopted in
relativistic field theories for the elimination of ultraviolet divergences
in perturbation expansions. For the anti-de Sitter group $SO(3,2)$ (for $P_5 =
1$) we obtained :
\be
 \Omega\left(p\right)=\frac{4 M^2}{4M^2-p^2} \;\; .
\label{10a}
\ee
	The analogous procedure starting from space-time $\M$ will provide
convergence factors $\Omega(x)$ which will eliminate infrared divergences. 
That conformally covariant theories may be free from divergences was also shown
by Mack and Todorov \cite{9}.

\section{The homogeneous space, action of conformal reflections.}

In reference \cite{8} it was shown how $\M_c$ and $\P_c$ may be represented as
homogeneous spaces of the conformal group 
$C = L \otimes D \rtimes P^{(4)}\rtimes  S^{(4)}$ 
(where $L$, $D$, $P^{(4)}$, $S^{(4)}$ stand for 
Lorentz-Dilatation-Poincar\'e-Special
conformal-transformations, respectively).

Precisely:
\be
 \M_c = \frac{C}{c_1} \;\; ; \;\;\;\;\; \P_c= \frac{C}{c_2} \;\; ,
\label{11}
\ee

where  $c_1 = L \otimes D \rtimes S^{(4)}$; $c_2 = L\otimes D \rtimes P^{(4)}$. $\M_c$ and $\M_c$ in eq.
(\ref{11}) are both isomorphic to $(S^3 \times S^1)/\Z_2$. Furthermore if we 
represent with $I$ a
conformal reflection (a reflection with respect to a plane orthogonal to
the $5^th$ or $6^th$ axis) then:
\be
 I \M_c I^{-1} = \P_c \;\; .
\label{12}
\ee

If we now consider the conformal group $C$ inclusive of reflections
(represented in $\R^{4,2}$ by $O(4,2)$), then $\M_c$ and $\P_c$ are two 
copies of the same
homogeneous space of $C$, transformed in each other by conformal reflections.
And then neither $\M_c$ nor $\P_c$ are automorphism spaces for $C$, but only 
the two
taken together; that is conformally compactified phase space. A conformal
reflection $I$ determines a convolution between $\M_c$ and $\P_c$ and then also
between $\M$ and $\P$. This duality, which could be named conformal duality, to
distinguished it from the quite different Fourier one, could be of
relevance for physics.

The action of $I$ in space-time $\M$, densely imbedded in $\M_c$ ,is well 
known; for $x_\mu \in \M$ : 
\be
 I \; : \;\; x_\mu \rightarrow I \left(x_\mu \right) = \pm \frac{x_\mu}{x^2} 
\;\; , \label{13}
\ee

which, for $x_\mu$ space like, is often interpreted as the map of every point,
inside a unit sphere $S^2$ in ordinary space\footnote{
 It is interesting to observe that, for $\M=\R^{2,1}$, for which the
conformal group is represented by the anti-de Sitter group $O(3,2)$, 
the sphere $S^2$ reduces to a
circle $S^1$ and then eq. (\ref{14}) reminds the Target Space duality
in string theory \cite{gvnvecc}, which then might be correlated with
conformal duality advocated in this paper.}, at a distance $x$
form its centre
to a point (on the same ray) at a distance $x^{-1}$:
\be
 I \; : \;\; x \rightarrow I\left(x\right)= \frac{1}{x} \;\; .
\label{14}
\ee

For the physical interpretation $x$ is thought to be dimensionless that is
represented by $x/L$ where $L$ is an arbitrary unit at length (and this breaks
conformal covariance which is already broken together with Lorentz
covariance in (\ref{14})). With this interpretation a micro  world 
($x/L \ll 1$) is transformed by $I$ to the macro world ($x/L \gg 1$) in 
ordinary 3D space.

The corresponding interpretation may be also adopted \cite{8} for the action of $I$ in
momentum space $\P = \R^{3,1}$ densely imbedded in $\P_c$, where, for 
$k_\mu \in  \P$:
\be
 I \; : \;\;\; k_\mu \rightarrow I\left(k_\mu\right) = \pm
\frac{k_\mu}{k^2}
\label{15}
\ee

and correspondingly
\be
 I \; : \;\; k \rightarrow I\left(k\right) = \frac{1}{k} \;\; .
\label{16}
\ee

If instead we take into account of (\ref{12}) eq. (\ref{14}) must be 
interpreted as bringing a point of $\M$ to a point of $\P$ and therefore we 
do not need to interpret $x$ as dimensionless, we may give it the meaning of 
a length and then eq. (\ref{14}) becomes:
\be
 I \; : \;\; x \rightarrow I\left(x\right)= \frac{1}{x} = k \in \P
\label{14a}
\ee

and similarly eq. (\ref{16}) becomes
\be
  I \; : \;\; k \rightarrow I\left(k\right)= \frac{1}{k} = x \in \M \;\; .
\label{16a}
\ee

Reminding that $C$ is an automorphism group for phase space and that the
physical momentum $p$ is obtained multiplying $k$ by an unit of action 
$H : p = H \cdot  k$, taking together the above equations we arrive to the 
following interpretation for the action of $I$ in physical phase space:
\be
  I \; : \;\; \frac{x p}{H} \rightarrow I\left(\frac{x p}{H}\right) = 
\frac{H}{x p} \;\; ;
\label{17}
\ee

which means: in phase space conformal inversion $I$ brings from regions of 
where the action is $\ll H$ to those where it is $\gg H$; that is from those
appropriate for the description of quantum phenomena (in the micro world) to those 
appropriate
for the description of classical phenomena (in the macro world). It could then 
represent a sort
of geometrical prerequisite for the realization of the correspondence
principle.

Now, since the conformal inversion $I$ brings also from $\M_c$ to $\P_c$ and
vice versa, it would appear that, since obviously $\M_c$ is appropriate for the
description of classical mechanics with the geometrical instrument of
euclidean geometry, momentum space could be the most appropriate for the
description of quantum mechanics, and in this space the most appropriate
geometrical instrument for its description seems to be spinor geometry.

\section{Spinor representation of quantum mechanics in momentum space.}

Fermions are the most elementary constituents of matter. Their properties
may be ideally described in the frame of spinor geometry, discovered by E.
Cartan \cite{2}, which, for what concerns us, may be summarized as follows
\cite{10}.

Given a real, $2n$ dimensional, vector space $V$ with scalar product $g$ with
signature $(k, l); k + l = 2n$, the corresponding Clifford algebra 
$\Cl(k,l)$, 
is central simple and has one, up to equivalence, representation:
\be
 \gamma \; : \;\; \Cl \left(k,l \right) = \rm{End} \;\; S_D
\label{18}
\ee
in a complex, $2^n$ dimensional space $S_D$ of Dirac spinors. 
If $\gamma_a$, obeying $ \left[ \gamma_a, \gamma_b\right]_+ = 2 g_{ab}$, are 
the generators of $\Cl(k, l)$ and $p_a$ the components of
a vector $p \in V$, a Dirac spinor $\psi$ may be defined through the Cartan's
equation
\be
 \sum_{a=1}^{2 n} p_a \gamma^a \psi =0.
\label{19}
\ee

For $\psi \not= 0$, we have that the vector $p$ is null: $p_ap^a = 0$ which 
implies that the directions of $p_a$ form the compact, projective quadric:
\be
 P_c = \frac{S^{k-1} \times S^{l-1}}{\Z_2} \;\; ,
\label{20}
\ee

\noindent 
of which eq. (\ref{3}) is a particular case for the signature $(4,2)$.

Eq. (\ref{19}) associates to each spinor $\psi $ a totally null plane in $V$ 
defined
by all null, mutually orthogonal vectors $p \in V$ satisfying it. 
When such a plane has dimension
$n$, that is maximal, the spinor $\psi$ was named simple by E. Cartan (and pure
by C. Chevalley).

E. Cartan has further shown how vectors of euclidean geometry in $V$ may be
conceived as bilinearly composed of spinors. 
In fact if we represent the generators $\gamma_a$  of $\Cl(k, l)$ as 
$2^n \times 2^n$ matrices acting on spinor space $S$, also the transposed 
matrices  $\gamma_a^t$, defined by
\be
  \gamma^t_a = B \gamma_a B^{-1},
\label{21}
\ee

will generate $\Cl(k, l)$ and $B$ is uniquely defined since $\Cl(k, l)$ is
simple, and they will act on the dual of $S_D$. We will have then, for 	
$\psi$ and $\phi \in S_D$ \cite{11}:
\be
 \psi \otimes B \phi = \frac{1}{2^n} \sum_{k=0}^{2n} 
 \gamma_{\mu_1} \gamma_{\mu_2} \dots \gamma_{\mu_k} 
   B_k^{\mu_1 \mu_2 \dots \mu_k} \left(\psi, \phi\right)
\label{22}
\ee

\noindent where
$$
  B_k^{\mu_1 \mu_2 \dots \mu_k} \left(\psi, \phi\right)=
 \langle B \psi , \gamma^{\mu_1} \gamma^{\mu_2} \dots \gamma^{\mu_k} 
 \phi \rangle \;\; ,
$$
 \noindent in which:
$$
 1 \leq \mu_1 < \mu_2 \dots \leq 2n.
$$

From (\ref{22}) we may then obtain:
\be
 \left(\gamma_a\psi \otimes B \phi \right) \gamma^a \psi = p_a \gamma^a \psi
\;\; , \label{23}
\ee

\noindent where
\be
 p_a = \langle B\phi, \gamma_a\psi \rangle.
\label{24}
\ee

Now $p_a p^a = 0$ for either $\phi$ or $\psi$ simple \cite{11} and then form 
(\ref{23}) we obtain identically Cartan's eq. (\ref{19}) where the $p_a$ are 
bilinearly expressed in terms of spinors.

Define now with $\gamma_{2n+1}$ the volume element of $\Cl(2n) = \Cl(k, l)$:
\be
 \gamma_{2n+1}=i^l \gamma_1  \gamma_2 \dots  \gamma_{2n} \;\; ,
\label{25}
\ee

it anti commutes with all the $\gamma_a$ , and it may be considered as the 
$(2n + 1)^{th}$ generator of $\Cl(k + 1, l)$ which is a non simple algebra 
while its even sub algebra $\Cl_0(2n + 1)$ is simple: $\Cl_0(2n + 1) = End S_P$
 and the associated spinors are named Pauli spinors.

The volume element $\gamma_{2n+1}$ defines the Weyl spinors 
$\psi_\pm$ of opposite helicity of $\Cl(2n)$ corresponding to each Dirac 
spinor $\psi$:
\be
 \psi_\pm = \frac{1}{2} \left(1 \pm \gamma_{2n+1} \right) \psi \; ; \;\;\;
 \psi_+ + \psi_- = \psi \;\; ,
\label{26}
\ee

building up the endomorphism spaces of the even subalgebra $\Cl_0(2n)$ of 
$\Cl(2n)$, which is non-simple.

For physical applications we need the vectors $p_a$ given in eq. (\ref{24}) to be
real. To this end we introduce the charge-conjugate spinor 
$\psi_c=C \bar{\psi}$  where $\bar{\psi}$ means $\psi$ complex conjugate and 
$C$ is defined by $C\gamma_a=\bar{\gamma_a}C$. Then we have  \cite{12}
that for the signature $(k, l) = (m + 1, m - 1)$ the vectors:
\be
 p_a^{\pm}=\langle B \psi_c , \gamma_a \left(1 \pm \gamma_{2n+1}
 \right) \psi \rangle \;\; ,
\label{25a}
\ee

are real (or imaginary) for $m$ even while complex for $m$ odd. That is
$p_a$
will be real for the signatures $(3,1)$, $(5,3)$, $(7,4) \dots$ 
(for $(4,2)$, that is for twistors, $p_a$ will be complex)\footnote{
It may be shown that they are also real for the lorenzian signature 
$(2n-1,1)$.}.

It is remarkable that in this way one obtains \cite{12} from Cartan's eq. (\ref{19}),
(where also the vectors $p_a$ are conceived as bilinearly composed by spinors)
not only the elementary equations of quantum mechanics in first
quantization,
however in momentum space; including Maxwell's equations (which somehow
constitute a bridge between quantum and classical physics), but also those
manifesting the so called internal symmetry.

In fact consider the following isomorphisms of algebras:

$\;\;\; \Cl(2n) \;\;\;$ is isomorphic to $\Cl_0(2n + 1) \;\;\;\;$ -- both simple 

$\;\;\; \Cl_0(2n) \;\;\;$ is isomorphic to $\Cl(2n + 1) \;\;\;\;$ -- both non-simple

which allows to consider a Dirac spinor associated with $\Cl(2n)$ as a
direct sum of Weyl spinor or of Pauli spinor which in turn may be
conceived as Dirac spinors of $\Cl(2n - 2)$ and so on. An elementary and
historical example is the space-time Dirac spinor, direct sum of right-
and left-handed Weyl spinors, which may also be considered as a doublet of
Pauli spinors associated with $\Cl(3)$ (for non relativistic motions). 

In this way from the Cartan's eq. (\ref{19}) for Weyl spinors associated
with $\Cl(5,3)$, taking into account of (\ref{25a}) the following equation
is obtained \cite{12}:  
\be
  \left( p_\mu \gamma^\mu \cdot {\mathbb I} + \vec{\pi} \cdot \vec{\sigma} \otimes \gamma_5+ m \cdot 
{\mathbb I} \right) N =0
\;\; , \label{26a}
\ee

where:
$ \vec{\pi}=\langle \tilde{N}, \vec{\sigma} \otimes \gamma_5 N \rangle$; 
 $N=\left[ \matrix{ \psi_1 \cr \psi_2 } \right]$; 
 $\tilde N=\left[ \tilde \psi_1 , \tilde \psi_2 \right]$; 
 $\tilde \psi_j=\psi_j^\dagger \gamma_0$, 

\noindent with  $\psi_1$,$\psi_2$ - space-time Dirac spinors, and 
$\vec{\sigma}=\left( \sigma_1,\; \sigma_2,\;\sigma_3 \right)$ Pauli matrices.

Eq. (\ref{26a}) is formally identical to the proton-neutron equation
interacting with the pseudoscalar isotriplet $\vec{\pi}$ representing the
pion, however in momentum space, and the internal isospin symmetry appears
as generated by the conformal reflections with respect to the planes
orthogonal to the $5^{th}$, $6^{th}$ and $7^{th}$ axis, and proton-neutron 
equivalence might represent a natural realization of quaternion algebra.

In fact it is known that a reflection with respect to a plane orthogonal to
$\gamma_a$ is represented in spinor space by $\psi \rightarrow \gamma_a \psi$;
 and, if $\gamma_a$ is time-like, it has to be 
substituted by $i\gamma_a$, if we impose that the square of a reflection 
equals the identity, which is the case 
for $\gamma_6$ in $\Cl(5,3)$, which brings to eq. (\ref{26a}). 
Furthermore also the
pseudoscalar nature of the pion triplet is uniquely determined from spinor
geometry since the representation in which the eighth component spinor $N$ is
a doublet of equivalent Dirac spinors, imposes for the gamma matrices to
have the form:
\be
 \Gamma_\mu = {\mathbb I} \otimes \gamma_\mu \;\; ; \;\; 
 \Gamma_5= \sigma_1 \otimes \gamma_5 
\;\; ; \;\; \Gamma_6=i \sigma_2 \otimes \gamma_5 \;\; ; \;\; \Gamma_7 = 
\sigma_3 \otimes \gamma_5 .
\label{27}
\ee

As we have seen the 8-component spinor $N$ of eq. (\ref{26a}) may be also 
considered as a doublet of Weyl spinors associated with $\Cl_0(4,2)$, or
twistors: $\Psi=\left[\matrix{\pi_+ \cr \pi_- } \right]$ obeying the 
Cartan's equation:
\be
 \left( p^a \tilde \Gamma_a +p^8 \cdot {\mathbb I} \right) \Psi =0 \;\; , 
\label{28}
\ee

where $\tilde \Gamma_a$ have the form:
\be
 \tilde\Gamma_\mu = \sigma_1 \otimes \gamma_\mu \;\; ; \;\; 
 \tilde\Gamma_5= \sigma_1 \otimes \gamma_5 \;\; ; \;\; 
 \tilde\Gamma_6=i \sigma_2 \otimes {{\mathbb I}} \;\; ; \;\; 
 \tilde\Gamma_7 = \sigma_3 \otimes {{\mathbb I}} .
\label{29}
\ee

Let us now define the $8 \times 8$ matrix
$$
 U = \left[ \matrix{ L & R \cr R & L}\right] = U^{-1} \;\; ,
$$
where $L=\frac{1}{2}\left(1+\gamma_5\right)$;  
$R=\frac{1}{2}\left(1-\gamma_5\right)$. It is easily seen that
$$
 U \Gamma_a U^{-1}= \tilde \Gamma_a \; ; \;\;\;\; U N = \Psi \; ; \;\;\;\;
 U \Psi = N \;\; ,
$$
from which we have that 
$(1 \otimes L) N= N_L \equiv (1 \otimes L) \Psi  = \Psi_L$.

But then eq.s (\ref{28}) and (\ref{26a}) may be summed to give\footnote{
The charged vector bosons derived from the $6-$vector $Z_a$, bilinearly
generated by twistors $\pi_+$, $\pi_-$, which are complex. They identically
satisfy the equation $Z_a \gamma^a \pi_+=0$ and 
$\bar{Z_a} \gamma^a \pi_-=0$, out of which the charged part of eq.
(\ref{30}) is obtained.}
\be
 \left( p_\mu + \vec A_\mu \cdot \vec \sigma  \right) \gamma^\mu N_L +
 B N_R = 0 \;\; ,
\label{30}
\ee

where $\vec A_\mu=\langle \tilde N, \vec \sigma \otimes \gamma_\mu N
\rangle$ 
and $N_R=\left(1 \otimes R \right)N$. 
If we now suppose that, of the two Dirac spinors
$\psi_1$,$\psi_2$, the first represents the electron: $e$ and the second the 
left handed neutrino: $\nu_L$ then eq. (\ref{30}) becomes:
\be
 \left( p_\mu + \vec A_\mu \cdot \vec \sigma  \right) \gamma^\mu
 \left[ \matrix{ e_L \cr \nu_L } \right]+ \alpha e_R = 0 \;\; , 
\label{31}
\ee

where $\alpha$ is a free parameter.

Eq. (\ref{31}) is the equation of the electroweak model, here derived from 
eq. s (\ref{28}) and (\ref{26a}) both obtained from eq. (\ref{19}) for 
$\Cl(5,3)$. It has been shown \cite{14}, \cite{15} that if one considers 
the triplet $e_L,e^c_L,\nu_L$ (or the corresponding
2-component Pauli spinors) to transform with $SU(3)$, the mixing angle 
$\Theta$ results determined such that $sin^2 \Theta = 0.25$.
Further details and consequences of these computations will be given 
elsewhere.

These examples, naturally derived from Cartan's eq. (\ref{19}), representing 
 some of the
basic equations of quantum physics in momentum space, in the frame of
spinor geometry, may induce to think that the method could be extended also
to higher dimensional spinor spaces, e.g. associated with $\Cl(9,1)$, in order to 
explain the multiplicities of elementary fermions (and bosons). In such spinor 
spaces the problem  of dimensional reduction (from 10 to 4,say) of
pseudo-euclidean vector spaces through {\it ad hoc} compactifications,
could be
avoided \cite{14} \cite{16}.

\section{Further aspects of conformal duality.}

Conformally compactified phase space, conceived as an automorphism space of
the extended conformal group, implies the conformal duality between
space-time and momentum space which appears as a convolution, determined by
conformal reflections. This duality might have two aspects of interest for
physics.

The first follows from its comparison with Fourier duality which is
defined trough functions, or physical fields, which may be defined in
space
time and its Fourier-dual momentum space. As we have seen in general, and
in particular in the soluble two dimensional case, for a compact phase
space such Fourier dual spaces may be only discrete and finite. As such
they might be named the "physical" spaces, to be distinguished from the
homogeneous or "mathematical" spaces $\M_c$ and $\P_c$ which are also 
finite but continuous. The "physical" discrete spaces will be both Fourier and
conformally-dual while the "mathematical" spaces will be only conformally
dual, and only the former should be the appropriate ones for description of
physical phenomena.

The second derives from the possible correlation of conformal duality with
the correspondence principle in so far it could be, as shown above, a sort
of geometrical prerequisite for the realization of the correspondence
principle, once one has found the motivation for identifying the 
unit of action H introduced in eq. (\ref{7}) with the Planck's
constant $\hbar$ (as in the de Broglie equality 
$p = \hbar \cdot k $). 
But it could perhaps, in any case, throw some light on some of the
still somehow mysterious aspects of the correspondence principle, as we will
try to show elsewhere.
Furthermore one could expect that some of the geometrical aspects, that is
of the topological and symmetry properties
 which are common to both $\M_c$ and $\P_c$
could be manifested by both classical and quantum physical systems independently
(and above) of the correlation, specifically predicated by the
correspondence principle (identification of wave functions with classical orbits for
high quantum
numbers). One of them is the $SO(4)$ symmetry which could be identified as
the maximal compact subgroup of $SO(4,2)$ and manifested by the presence of
its isometry sphere $S^3$ in both $\M_c$ and in its conformally dual $\P_c$; 
and then
to be expected in both classical and quantum mechanical stationary (non
relativistic) systems in ordinary- and momentum-space respectively. 
These  systems exist they are the planetary motions
in space-time and the H-atom in momentum space. In fact it is remarkable
that the $SO(4)$ symmetry of the H-atom was discovered  by V. Fock in the $S^3$
compactification of momentum space \cite{17}, which suggests that this $SO(4)$
symmetry might be a consequence of conformal duality, rather than being  ``accidental'', 
as it was named by W. Pauli when discovered.

With R. Raczka \cite{18} we have also conjectured an eigenvibration of the
$S^3$
sphere
of the Robertson Walker universe represented by
\be
M_{RW} = S^3 \times R^1,
\label{32}
\ee
(which is often considered as a natural realization of $\M_c$ given by 
eq. (\ref{1})
where $R^1$ is the infinite covering of $S^1$), in order to explain a 
remarkable regularity in the distribution of distant galaxies 
\cite{19} (more than eleven
peaks equally spaced by about $4 \cdot 10^8$ light years, in the direction
of the
North and South galactic poles). We have shown \cite{20} how the
astronomical data
are well represented by the most symmetric spherical harmonic of $S^3$:
\be
 Y_{n,0,0} = k_n \frac{\rm{sin} \; \left(n+1\right) \chi }{\rm{sin} \; \chi} 
\label{33}
\ee
where $\chi$ is the geodesic distance from center of the eigenvibration.

If further astronomical observations will confirm this model it would
represent a remarkable test  of conformal duality since eq.
(\ref{33}) is
exactly the eigenfunction of the stationary S-states of the H-atoms found
by V. Fock precisely in $S^3$ compactification of momentum space (eq. (26) of ref. 
\cite{17}).  Then the Universe and the
H-atom would constitute an example of realization of conformal duality,
representing two conformally dual systems having the
same eigenfunction : the first in ordinary space and the second in the conformally 
dual momentum space.

\end{document}